# Efficient demodulation scheme for multilevel modulation based optical camera communication

ZIHAO ZHOU[1], MAOLIN LI[2] AND WEIPENG GUAN[3]

[1] *School of Electronic and Information Engineering, South China University of Technology, Guangzhou , Guangdong 510640, China*
[2] *School of Automation Science and Engineering, South China University of Technology, Guangzhou 510640, China*
[3] *School of Automation Science and Engineering, South China University of Technology, Guangzhou 510640, China*

**Abstract:**

## 1. INTRODUCTION

As a short-range wireless communication technology, visible light communication (VLC) has attracted considerable interest due to its better security, lower time delay and higher bandwidth [1]. Most existing VLC systems use the positive-intrinsic-negative (PIN) and the avalanche photodiode (APD) as receiver (Rx) [2]. However, they are not available in commercial off-the-shelf devices, which reduces the flexibility and universality of VLC systems to a certain extent [3]. In recent years, thanks to the great leap with regard to price and performance of image sensor technology, even the least expensive smartphones nowadays are equipped with high-resolution complementary metal-oxide semiconductor (CMOS) cameras. Hence, this motivates using CMOS camera for the VLC systems, which is known as optical camera communication (OCC).

Utilizing the rolling shutter effect (RSE) of CMOS camera is a mainstream method to improve data rate in OCC system[4,5]. The camera will exposure line by line subsequently, hence when the transmitting rate is higher than the frame rate, the bright and dark stripes can be observed in each received frame, which means that more information can be carried in one image frame [6]. An interesting packet loss compensation method based on RSE is also proposed in [7]. And based on RSE, many technologies can also play its real power, such as wavelength division multiplexing (WDM) [8], decoding scheme based on fourier transform [9], more advanced modulation techniques like rolling OFDM [10] and modulation format shifting[2]. However, for most binary OCC systems[11-17], as the stripe width becomes narrower, the bit-error-rate (BER) performance of the system will deteriorate dramatically which limits the rate improvement. Compared with binary RSE-based OCC system, multilevel modulation can provide higher data rate with the same fringe width. In [18,19], pulse width modulation (PWM) is adapted to realize multilevel modulation indirectly. In [20], Shi, J. et al proposed a multilevel modulation scheme based on the combination of on-off-keying (OOK) and Manchester code, but the scheme needs to decode by the difference between pixels, which makes the whole system particularly sensitive to noise and brings high BER. It is worth noting that the decoding algorithm proposed in [20] is customized according to the coding structure and it is not suitable for [18,19]. And the blur curved fitting method proposed in [19] is difficult to obtain accurate results when the distance increases. To sum up, there is a lack of a general and effective threshold algorithm in the research of multilevel-based OCC.

Another obstacle that limits multilevel-based OCC technology is the sampling point offset (SPO). Although the sampling principle of multilevel based OCC system is consistent with that of binary OCC system, as the order of the system increases, the interval between symbols will become narrower, which puts forward high requirements for the accuracy of sampling. Conventional clock recovery (CR) based on synchronization head is widely used in binary OCC systems[11-17,21]. After the received data is processed with threshold scheme, sampling is required to extract the binary signal. And generally the sampling interval is estimated with synchronization head. However, the tolerance of high-order OCC system to sampling offset is reduced, and the sampling offset of a symbol will affect the subsequent sampling of a series of symbols. But in the previous works on multilevel modulation [18-21], the issue of SPO, which significantly affects the transmission performance, has not been taken into account sufficiently.

In this paper, firstly, a new hybrid code structure based on the overlapping of two light sources to produce the effect of multi-voltage amplitudes is proposed and experimentally demonstrated. And we also proposed an efficient polarity reversal threshold (RRT) algorithm for multilevel based OCC system. Then taking the issue of SPO into account, a novel adaptive sampling method (ASM) is proposed, which can effectively alleviate the problem of SPO and further enhance the performance of multilevel OCC. It is demonstrated that a data rate of 8.4 Kbit/s can be achieved by applying the proposed two algorithms.

## 2. THEORY

### A. Multilevel modulation scheme

Generally, multilevel modulation requires multi-voltage amplitudes to generate different intensity light, which undoubtedly increases the complexity of transmitter circuit. Inspired by ref [20], we designed a new hybrid code structure based on the overlapping of two light sources to produce the effect of multi-voltage amplitudes. But different from the ref [20], LED1 and LED2 are modulated by non-return-to-zero (NRZ)-OOK and return-to-zero (RZ)-OOK signal respectively, which is shown in Fig.1(a). NRZ-OOK modulation is used to deliver binary data by changing "on" and "off" state of the light. It is a simple way that using "on" state of light denotes symbol "1" and "off" state of light denotes symbol "0". Similarly, the RZ-OOK does not send a pulse when it transmits symbol "0", but instead sends a pulse with a duty cycle less than 100% (typically 50%) when it transmits symbol "1". The following will further elaborate the significance of this improvement. We record the pulse width of NRZ-OOK as $T_1$ and the pulse width of RZ-OOK and Manchester as $T_2$. Due to the adoption of RSE, when $T_2$ is large enough, that is, the CMOS sensor can capture stripes of a pulse with a width of $T_2$, both the scheme in this paper and the scheme in [20] will eventually generate independent waveforms of three symbols: 0(00B), 1(01B) and 2(10B), which will reflect three stripes of black, gray and white on the image, as shown in Fig.1(c). But this leads to an incomplete representation of the code, as the system cannot produce symbol 3(11B).

On the other hand, as we all know that the pulse width cannot be reduced indefinitely in order to ensure the resolution of the fringe in the existing OCC system. As shown in Fig.1(d), when $T_2$ is reduced to the point where the CMOS sensor cannot capture $T_2$-width pulses but can only capture $T_1$-width pulses, the $T_1$-width pulse is still reflected in the image as black and white stripes, while the function of the $T_2$-width pulse at this time is to superimpose two different gray average values on the basis of the black and white stripes. Hence the hybrid coding of NRZ-OOK and RZ-OOK can generate four kinds of multilevel signals with different mean values. However, in [20], due to the symmetry of Manchester coding, it is difficult to generate enough signals with different mean values within the same bit duration. Fig. 1(b) illustrates the difference between this paper and [20].

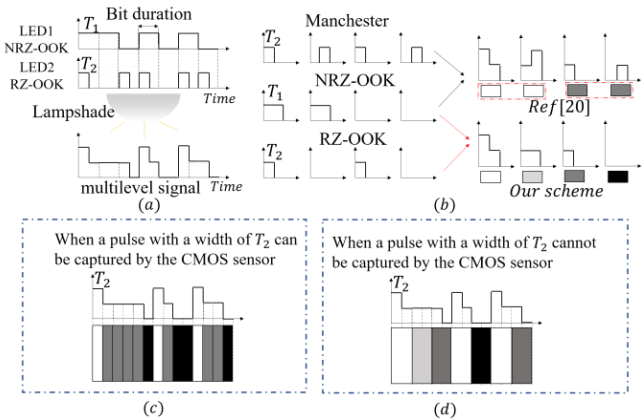

Fig.1. (a) Principle of multilevel modulation proposed in this paper. (b) The difference between this paper and [20] in symbol waveform. (c) Situation when a pulse with a width of $T_2$ can be captured by the CMOS sensor. (d) Situation when a pulse with a width of $T_2$ cannot be captured by the CMOS sensor.

Therefore, in the 4-ary OCC system discussed above, the optical signal waveform has the following form:

$$s(t) = \sum_{n=-\infty}^{+\infty} [a_n g_T(t - nT) + b_n g_T(2t - nT)] \quad (1)$$

Where $a_n$ and $b_n$ are the amplitudes of the $n$-th waveform of signals transmitted by LED1 and LED2 respectively, and their values are discrete 0 or 1. $g_T(t)$ is a pulse signal with a duration of one symbol period $T$. Fig. 2a-b shows the difference of average grayscale value between the four symbols 3, 2, 1 and 0. Fig. 2c-d shows the effect of sending four combinations of waveforms using the scheme [20] when a pulse with a width of $T_2$ cannot be captured by the CMOS sensor. It is worth noting that in reality due to the limitation of LED response speed, there will be a transition time from one brightness to another, and the greater the brightness difference, the longer the transition time required.

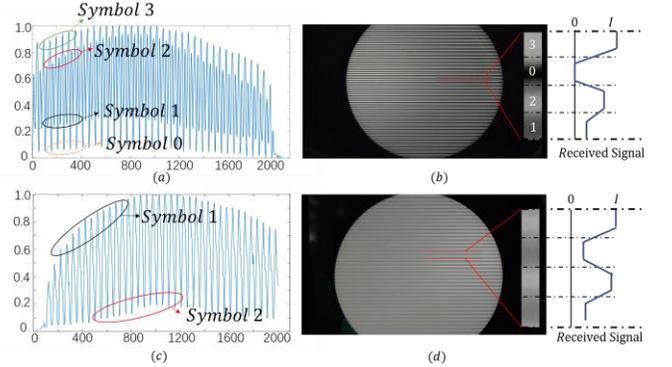

Fig.2. The difference of average gray value between different symbols.

### B. Polarity reversal threshold algorithm (PRT)

At the receiver, we adopted the algorithm proposed by our previous work [5] to select the appropriate column matrix from the frame image captured by CMOS sensor, and record it as $p(i)$. Next, thresholding is needed to restore the digital logic. Most of the existing thresholding methods, such as iterative thresholding [12], are only suitable for binary OCC system. Therefore, we proposed an effective polarity reversal thresholding (PRT) algorithm for multilevel based OCC system in this paper. Flow diagram in Fig.3 showing the working principle of PRT algorithm. Suppose that there is an M-ary OCC system (M = 2, 4, 8, etc). Firstly, by applying third-order polynomial fitting to the selected column of grayscale values $p(i)$, (i=1,2,…1080), the thresholding curve referred to as middle threshold $Th_m(i)$ is generated, which is shown by the red line in Fig.4(a). Next, enter the first judgment, if (M-4)/2 is less than 0 then exit the algorithm. It is equivalent to using the third-order polynomial fitting to thresholding a binary OCC system. Otherwise the next step is to reverse the polarity, we update the original column pixels $p(i)$ according to Eq.(2) to get a new set of values $p_1(i)$:

$$p_1(i) = p(i) < Th_m(i)? [2Th_m(i) - p(i)] : p(i) \quad (2)$$

The gray value distribution of $p_1(i)$ is shown in Fig.4(b). Among them, the expression $2Th_m(i) - p(i)$ represents the process of low grayscale values inversion with $Th_m(i)$ as the axis. Then apply third-order polynomial fitting to the $p_1(i)$ to obtain a new thresholding curve referred to as upper threshold $Th_h(i)$, as shown by the red line in Fig.4(c). In parallel, we make another update to the original column pixel $p(i)$ according to Eq.(3). Let grayscale values in $p(i)$ greater than thresholding values reverse with $Th_m(i)$ as the symmetry axis, the update formula of gray value is:

$$p_2(i) = p(i) < Th_m(i) ? p(i) : [2Th_m(i) - p(i)] \quad (3)$$

The gray value distribution of $p_2(i)$ is shown in Fig.4(e). And a new lower threshold $Th_l(i)$ in Fig.4(f) is obtained by applying third-order polynomial fitting to the $p_2(i)$. In order to eliminate the noise to the threshold interval division, we need to refine the threshold $Th_h(i)$ and $Th_l(i)$. Update the grayscale values by adopting the following two formulas respectively:

$$p_3(i) = p(i) < Th_h(i) ? Th_h(i) : p(i) \quad (4)$$
$$p_4(i) = p(i) < Th_l(i) ? p(i) : Th_l(i) \quad (5)$$

As shown in Fig.4(d) and (g), the final refined threshold curve termed as $Th_{high}(i)$ and $Th_{low}(i)$ can be obtained by applying third-order polynomial fitting to $p_3(i)$ and $p_4(i)$. After completing the above steps, three different thresholds are generated. This means that for the above 4-ary OCC system, we have completed the whole process of PRT algorithm. For higher order OCC systems (M > 4), the remaining thresholds can be calculated by the following formula:

$$\begin{cases} Th_m + m \times \dfrac{Th_{high} - Th_m}{\dfrac{M-2}{2}} \\ Th_m - m \times \dfrac{Th_m - Th_{low}}{\dfrac{M-2}{2}} \end{cases} \quad m = 1,2,\cdots,\dfrac{M-4}{2} \quad (6)$$

Fig. 4 shows the effect of applying PRT algorithm for a multilevel based OCC system when M = 4. Fig. 4(h) illustrates that the normalized gray range can be divided into four intervals corresponding to the symbols 0, 1, 2 and 3 respectively through the above $Th_m, Th_{high}$ and $Th_{low}$.

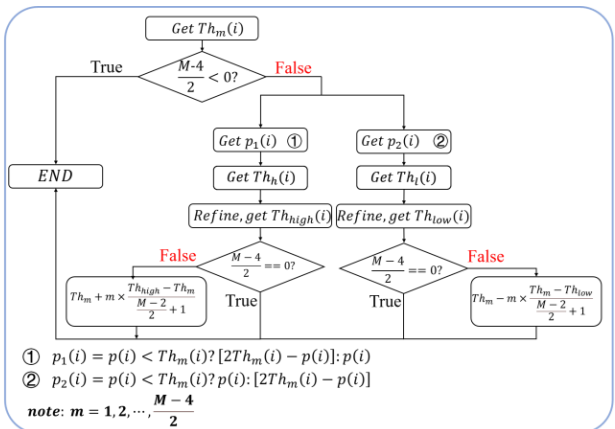

Fig.3. Flow diagram showing the working principle of PRT scheme.

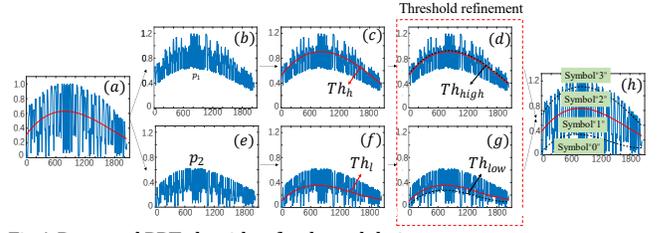

Fig.4. Proposed PRT algorithm for demodulation.

### C. Adaptive sampling method (ASM)

The key steps of ASM include finding the auxiliary extreme point, length transformation and sampling. Firstly, linear interpolation is used to make the fringe width of one bit be odd and recorded as $X$. Hence we can obtain an array $D$ that contains a positive integer multiple of $X$ where $D = [X, 2X, 3X, \cdots]$. The length of $D$ can be determined according to the coding method, and $N$ is used to denote the length of $D$. Next step is to find the location of all the local maximum and minimum in the column matrix to form arrays $L_{max}$ and $L_{min}$ with length of $N_1$ and $N_2$ respectively. All the elements of $L_{max}$ and $L_{min}$ are merged and sorted to get the matrix $L$ with length of $N_1 + N_2$. Then we can get the distance between the extreme points using first order difference as $CL(n) = L(n+1) - L(n)$, $(n = 1,2, \cdots N_1 + N_2 - 1)$. The elements in the $CL$ represent the distance between adjacent extremes. In order to determine effective auxiliary extremum and eliminate the interference of noise, find out the elements in the interval $\left[\left\lceil\dfrac{X}{2}\right\rceil, \left\lfloor\dfrac{3X}{2}\right\rfloor\right]$ in the vector $CL$ and record the subscript as $i$, where $\lceil\cdot\rceil$ is rounding up symbol and $\lfloor\cdot\rfloor$ is rounding down symbol. Then we take $L(i), L(i+1)$ as a set of auxiliary extreme points. We combine all the found auxiliary extreme points into a matrix $FL$. $N_{FL}$ is used to denote the length of $FL$.

Next comes the second step: re-interpolation. Calculate the first order difference for the array $FL$ as $delta(j) = FL(j+1) - FL(j)$ $(j = 1,2, \cdots N_{FL} - 1)$. And construct the following objective function:

$$f(k) = \big(delta(j) - D(k)\big)^2 \quad (7)$$

Then we can find $k$ which makes $f(k)$ minimum, where $k \in (1, N)$. Where $k$ means how many bits should theoretically be contained in the pixel matrix of $delta(j)$ length. As we mentioned above, $X$ pixel rows represent one bit after interpolation, hence theoretically the number of pixel rows occupied by $k$ bits should be $k \times X$. This sequence of length $delta(j)$ needs to be transformed (interpolated or extracted) into a sequence of length $k \times X$, so as to facilitate accurate sampling in the following step. The ratio of sequence length can be calculated by Eq.(8).

$$Ratio = \dfrac{k \times X}{delta(j)} \quad (8)$$

This transformation can be divided into three cases:

• $Ratio > 1$: It shows that the length of $delta(j)$ sequence needs to be expanded, hence spherical linear interpolation will be adopted to make the length of this sequence become $k \times X$. At the same time, we need to replace the last element of the interpolated sequence with the (j+1)-th auxiliary extremum point. Fig.5 shows the processing method when $Ratio > 1$.

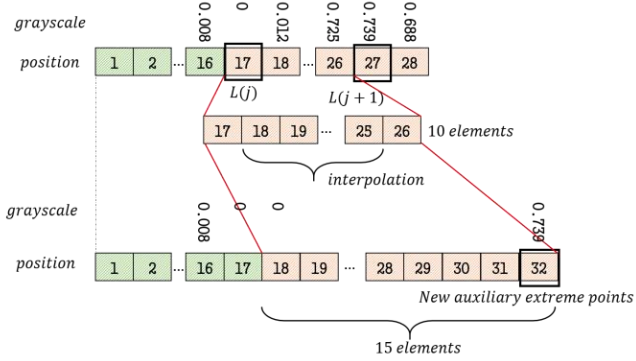

Fig.5 Processing mechanism when $Ratio > 1$

• $Ratio < 1$: It means that some values need to be pulled out of the sequence randomly to make the length of this sequence become $k \times X$. Similarly, the last element of the interpolated sequence should be replaced with the (j+1)-th auxiliary extremum point. And the processing method when $Ratio > 1$ is illustrated in Fig.6.

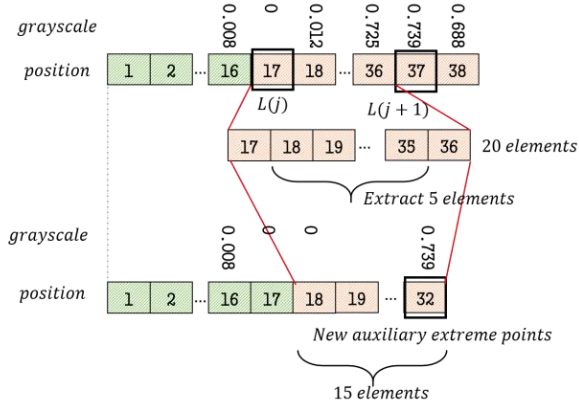

Fig.6 Processing mechanism when $Ratio < 1$

• $Ratio = 1$: When $Ratio$ is equal to 1, it means that there is nothing to do but replace the last element of the sequence with (j+1)-th auxiliary extremum point hence it is already has a standard length.

After the above series of operations, a new set of auxiliary extremum points are obtained, and record it as $FL_{new}(t), (t = 1, 2, \cdots, N_{FL} - 1)$. They are one less than the old auxiliary extremum points (that is, the first extremum point which is discarded in the algorithm). There is no change in the gray value, but there will be a displacement in the position. The last step is to sample according to the auxiliary extremum.

The sampling points between the two auxiliary extreme points can be obtained according to the following formula:
$$FL_{new}(m) + \beta \times X \quad (9)$$
Where $\begin{cases} \beta = 0, \cdots, \left(\frac{FL_{new}(m+1) - FL_{new}(m)}{X} - 1\right) \\ m = 1, 2, \cdots N_{FL} - 2 \end{cases}$. It is worth noting that $FL_{new}(m)$ is a multiple of $X$ due to the length adjustment in the previous step. In addition, all auxiliary extreme points need to be taken as sampling points.

Fig. 7 shows the comparison of the sampling results between ASM and traditional CR method. It can be found that there are errors even in header positioning when sampling using the CR method.

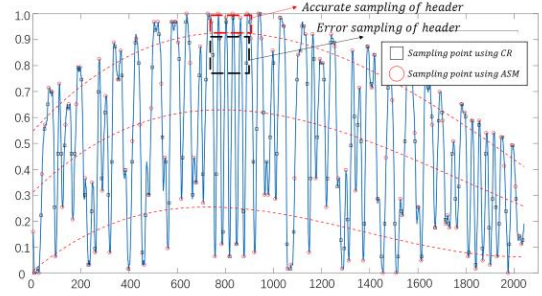

Fig.7. Comparison of sampling results between ASM and traditional CR methods

## 3. Experimental Section

### A. Experimental Setup

Fig. 8(a) shows the experimental setup of the quaternary OCC system (M=4) using mobile phone camera as Rx. Two sets of data modulated by the NRZ-OOK and the RZ-OOK are offline generated from personal computer (PC) and loaded into the single port read-only memory (ROM) of field-programmable gate array (FPGA) from Altera Cyclone Ⅳ series. The general-purpose input/output (GPIO) pin outputs high and low voltage, which correspond to the data bit "1" and "0" stored in ROM respectively. Due to the restricted current drive of GPIO, a LED Driver is placed after the FPGA to increase the current drive and control the brightness of white-light LED with color temperature of 5000 K. Therefore, the GPIO pin of FPGA is responsible for controlling the LED driver rather than directly driving the LED. The LED1 and the LED2 are modulated by the NRZ-OOK and the RZ-OOK signal respectively. Subsequently, two sets of data are separately emitted by the LEDs (CREE-XML2). A LED light diffuser is used to generate the hybrid signal including the NRZ-OOK and the RZ-OOK signal.

In the case of RZ-OOK coding, symbol "1" stands for 1/2 bit. Due to the same sampling clock for both GPIO1 and GPIO2 ports, each bit of NRZ-OOK signal sends two times consecutively at bit period. The structure of data packet contains a 10-bit header "1010101010" and payload. Each packet is transmitted three times to mitigate the gap-time effect and guarantee each image frame captured by the mobile-phone contains a complete data packet [16].

Through a 20cm atmospheric channel, at the receiver, a plano-convex lens is adopted in front of a smartphone (Huawei P20 Pro with resolution of 1920×1080 and frame rate of 60 fps). The function of the plano-convex lens here is to condense the incoming light and increase the imaging area of the LED so as to improve the communication distance while maintaining the data rate. The smartphone camera is employed in a video capture mode with a manual configuration of the exposure compensation of -4 EV, an ISO value of 1000 and the exposure time is adjusted to 250 $\mu s$. The key hardware parameters are shown in Table 1.

Table 1. Hardware parameters of the experiment

| Parameter | Value |
| --- | --- |
| The focal length of lens/mm | 49mm |
| The resolution of the camera | 1920×1080 |
| The ISO of the camera | 1000 |
| Frame rate/fps | 60 |
| Exposure compensation/EV | -4 |
| Exposure time/$\mu s$ | 250 |
| The diameter of the LED /cm | 4.5 |
| The power of the LED /W | 9 |
| Voltage of the LED/V | 3.2 |
| Current of the LED/mA | 22 |
| Diameter of the LED/cm | 3 |

In the demodulation part, firstly, the raw movie file captured by the mobile-phone CMOS image sensor is converted into grayscale format. And a proper column matrix of grayscale values is selected and normalized from the grayscale image. Linearly interpolated is adopted to increase the effective sampling points. Meanwhile, histogram equalization is carried out to increase the extinction ratio (ER). Next, the polarity reversal threshold (PRT) algorithm proposed in this paper is used to define the different symbols. To mitigate the SPO, we also proposed an adaptive sampling method (ASM). When sampling is completed, header needs to be located in the image frame for packet reconstruction. After the above steps, recovered data packet is examined for BER calculation, so as to evaluate the transmission performance of the multilevel-based OCC system. The experimental scene is illustrated in Fig.8(b).

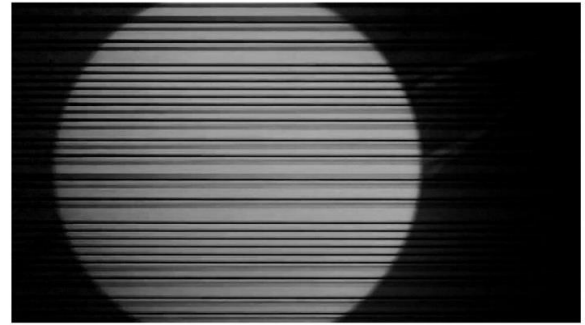

Fig.9 The red channel of one of the frames captured by rhe CMOS image sensor.

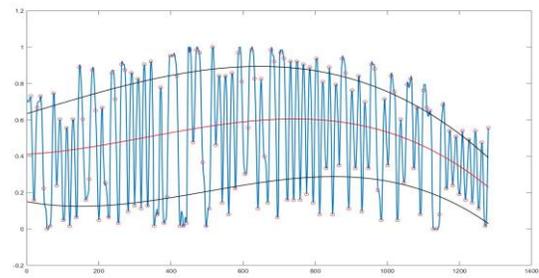

Fig.10 The grayscale curve after normalization and the sampling results.

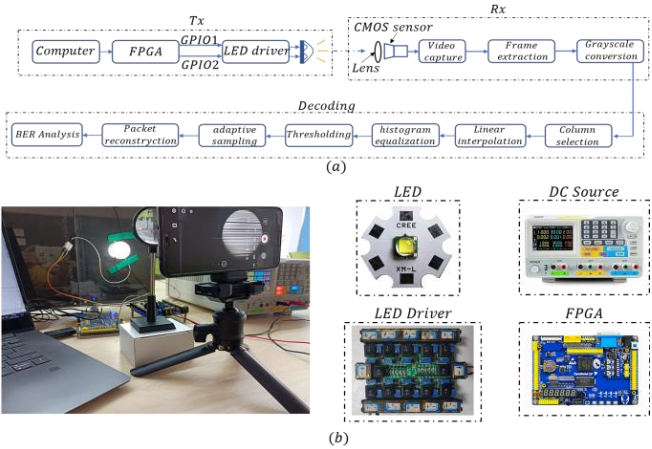

Fig.8(a) Block diagram of the experimental setup of the multilevel-based OCC system. (b) Experimental scene.

### C. Results and Discussion

The experimental results using our scheme are in line with our expectations. We further tested with two RGB LED lights.However, we used NRZ_OOK and PWM(Paulse width modulation) to modulated the two sets of data for each colour.The structure of data packet are the same as above.We separated each image frame to three channels and usedd our scheme to decode.Fig.9 shows the red channel of one of the frames. The demodulation results are shown in Fig.10.

## 4. Conclusion

**Funding.** This work was supported by Research and Development Program in Key Areas of Guangdong Province (2019B010116002); National Undergraduate Innovation and Entrepreneurship Training Program (202010561020); Guangdong Science and Technology Project under Grant (2017B010114001).

**Disclosures.** The authors declare no conflicts of interest.

**Acknowledgments.** The authors thanks the Prof.Wen and Prof. Chen, for providing necessary financial assistance.

**Data availability.** No data were generated or analyzed in the presented research.

## References

1. Boubezari, R., Le Minh, H., Ghassemlooy, Z., & Bouridane, A. "Smartphone camera based visible light communication". Journal of Lightwave Technology, **34**, 4121-4127.(2016)
2. Shi, J., He, J., Jiang, Z., & Chang, G. K. "Modulation format shifting scheme for optical camera communication". IEEE Photonics Technology Letters, **32**, 1-1, (2020).
3. V Matus, VG Yánez, Zvanovec, S., Rabadan, J., & R Pérez-Jiménez. "Sandstorm effect on experimental optical camera communication". Applied Optics, **60**, 75-82 (2021).
4. Le, T., Le, N. T., & Jang, Y. M. "Performance of rolling shutter and global shutter camera in optical camera communications", in International Conference on Information & Communication Technology Convergence. IEEE. (2015), pp. 124-128.
5. Zhou, Z.; Wen, S.; Li, Y.; Xu, W.; Chen, Z.; Guan, W. "Performance Enhancement Scheme for RSE-Based Underwater Optical Camera Communication Using De-Bubble Algorithm and Binary Fringe Correction". Electronics, **10**, 950. (2021).
6. Chow, C. W., Chen, C. Y., & Chen, S. H. "Visible light communication using mobile-phone camera with data rate higher than frame rate". Optics Express, **23**, 26080-26085, (2015).
7. Liqiong, Liu, Rui, Deng, Lian-Kuan, & Chen. "47-kbit/s rgb-led-based optical camera communication based on 2d-cnn and xor-based data loss compensation". Optics express, **27**,33840-33846, (2019).
8. D Rui, Jing, H., Yang, H., Jin, S., & Lin, C. "2.38 kbits/frame wdm transmission over a cvlc system with sampling reconstruction for sfo mitigation". Optics Express, **25**, 30575-30581, (2017).
9. Xiao, Y., Guan, W., Wen, S., Li, J., Li, Z., & Liu, M. The Optical Bar Code Detection Method Based on Optical Camera Communication Using Discrete Fourier Transform. IEEE Access, **8**, 123238-123252, (2020)
10. Nguyen, H., Thieu, M. D., Nguyen, T., & Jang, Y. M. "Rolling OFDM for image sensor based optical wireless communication". IEEE Photonics Journal, **11**, 1-17, (2019)
11. Liang, K., Chow, C. W., Yang, L., & Yeh, C. H. "Thresholding schemes for visible light communications with cmos camera using entropy-based algorithms". Optics Express, **24**, 25641. (2016)
12. Liu, Y., Chow, C. W., Liang, K., Chen, H. Y., Hsu, C. W., Chen, C. Y., & Chen, S.H. "Comparison of thresholding schemes for visible light communication using mobile-phone image sensor". Optics express, **24**, 1973-8. (2016)
13. Guan, W., Wu, Y., Xie, C., Fang, L., Liu, X., & Chen, Y. "Performance analysis and enhancement for visible light communication using cmos sensors". Optics Communications, **410**, 531-545. (2018)
14. Chen, C. W., Chi-Wai, C., Yang, L., & Chien-Hung, Y. "Efficient demodulation scheme for rolling-shutter-patterning of cmos image sensor based visible light communications". Optics Express, **25**, 24362. (2017)
15. Do, T. H., & Yoo, M. "Performance analysis of visible light communication using cmos sensors". The Journal of Korean Institute of Communications and Information Sciences, **40**, 309. (2016)
16. Chow, C. W., Chen, C. Y., & Chen, S. H. "Enhancement of signal performance in led visible light communications using mobile phone camera". Photonics Journal IEEE, **7**, 1-7. (2015).
17. Zhang, Z., Zhang, T., Zhou, J., Qiao, Y., Yang, A., & Lu, Y. "Performance enhancement scheme for mobile-phone based vlc using moving exponent average algorithm". IEEE Photonics Journal, **9**, 1-7. (2017)
18. Kinoshita, M., Zinda, T., & Chujo, W. "Multilevel Modulation by LED Luminance Distribution for Optical Camera Communication", in 2018 International Symposium on Antennas and Propagation (ISAP). IEEE, pp. 1-2
19. Lee, J. W., Yang, S. H., & Han, S. K. "Optical pulse width modulated multilevel transmission in cis based vlc". IEEE Photonics Technology Letters, **29**, 1257-1260, (2017).
20. Shi, J., He, J., Deng, R., Wei, Y., Long, F., Cheng, Y., & Chen, L. "Multilevel modulation scheme using the overlapping of two light sources for visible light communication with mobile phone camera", Optics Express. **25**, 15905-15912, (2017).
21. Chen, C. W., Chi-Wai, C., Yang, L., & Chien-Hung, Y. "Efficient demodulation scheme for rolling-shutter-patterning of cmos image sensor based visible light communications". Optics Express, **25**, 24362. (2017).
22. Guan W, Huang L, Wen S, et al. Robot Localization and Navigation Using Visible Light Positioning and SLAM Fusion[J]. Journal of Lightwave Technology, 2021, 39(22): 7040-7051.
23. Huang L, Wen S, Yan Z, et al. Single LED positioning scheme based on angle sensors in robotics[J]. Applied Optics, 2021, 60(21): 6275-6287.
24. Yan Z, Guan W, Wen S, et al. Multi-robot Cooperative Localization based on Visible Light Positioning and Odometer[J]. IEEE Transactions on Instrumentation and Measurement, 2021.
25. Song H, Wen S, Yang C, et al. Universal and Effective Decoding Scheme for Visible Light Positioning Based on Optical Camera Communication[J]. Electronics, 2021, 10(16): 1925.
26. Guan W, Chen S, Wen S, et al. High-accuracy robot indoor localization scheme based on robot operating system using visible light positioning[J]. IEEE Photonics Journal, 2020, 12(2): 1-16.
27. Guan W, Huang L, Hussain B, et al. Robust robotic localization using visible light positioning and inertial fusion[J]. IEEE Sensors Journal, 2021.